\documentclass[10pt,journal,compsoc]{IEEEtran}

\usepackage[justification=centering]{caption}
\usepackage{algorithm,algpseudocode}
\usepackage{amsmath}
\usepackage{graphicx}
\usepackage{times}                               
\usepackage{graphicx}
\usepackage{tabularx}
\usepackage{array}
\usepackage{amsmath}
\usepackage{amssymb}
\usepackage{units}
\usepackage{textcomp,xspace}
\usepackage{enumerate}
\usepackage{multirow}
\usepackage{multicol}
\usepackage{subfigure}
\usepackage{color}

\makeatletter
\def\BState{\State\hskip-\ALG@thistlm}
\makeatother

\ifCLASSINFOpdf

\else

\fi

\begin{document}

\title{Accelerating BLAS and LAPACK via Efficient Floating Point Architecture Design}

\author{Farhad Merchant, Anupam Chattopadhyay, ~\IEEEmembership{Senior Member, IEEE,} Soumyendu Raha, \\S K Nandy, ~\IEEEmembership{Senior Member, IEEE,} and Ranjani Narayan
\IEEEcompsocitemizethanks{\IEEEcompsocthanksitem Farhad Merchant and Anupam Chattopadhyay are with School of Computer Science and Engineering,
Nanyang Technological University, Singapore\protect\\
E-mail: \{mamirali,anupam\}@ntu.edu.sg
\IEEEcompsocthanksitem Soumyendu Raha and S K nandy are with Indian Institute of Science, Bangalore \IEEEcompsocthanksitem Ranjani Narayan is with Morphing Machines Pvt. LTd. }
\thanks{Manuscript received October 20, 2016;}}

\IEEEtitleabstractindextext{%
\begin{abstract}
Basic Linear Algebra Subprograms (BLAS) and Linear Algebra Package (LAPACK) form basic building blocks for several High Performance Computing (HPC) applications and hence dictate performance of the HPC applications. Performance in such tuned packages is attained through tuning of several algorithmic and architectural parameters such as number of parallel operations in the Directed Acyclic Graph of the BLAS/LAPACK routines, sizes of the memories in the memory hierarchy of the underlying platform, bandwidth of the memory, and structure of the compute resources in the underlying platform. In this paper, we closely investigate the impact of the Floating Point Unit (FPU) micro-architecture for performance tuning of BLAS and LAPACK. We present theoretical analysis for pipeline depth of different floating point operations like multiplier, adder, square root, and divider followed by characterization of BLAS and LAPACK to determine several parameters required in the theoretical framework for deciding optimum pipeline depth of the floating operations. A simple design of a Processing Element (PE) is presented and shown that the PE outperforms the most recent custom realizations of BLAS and LAPACK by $1.1$X to $1.5$X in GFlops/W, and $1.9$X to $2.1$X in Gflops/$mm^2$.  


\end{abstract}

}

\maketitle

\IEEEpeerreviewmaketitle

\section{Introduction}
Domain specific computing platforms have gained immense popularity in the last decade. For domain specific computing, custom architectures are developed for efficient realization of several algorithms/computations pertaining to the domain of interest. Several architectural parameter such as size of the memory, bandwidth in the memory subsystem, and compute resource choices are chosen that are specific to the domain of interest. For domain specific computing, accelerators are preferred as an ideal underlying platform due to their better power performance over general purpose computers \cite{mudge3}\cite{lac1}\cite{baluni1}. While accelerators like General Purpose Graphic Processing Units (GPGPUs) dissipate more power than desired, there are several domain specific accelerators designed to overcome this shortcoming of GPGPUs \cite{lac1}\cite{hyper1}.

Domain customized platforms and/or accelerators are gaining popularity due to their area and power performance \cite{lac3}\cite{Gregg2}\cite{cgr2}\cite{cgr3}. Performance in these accelerators is achieved by setting several architectural parameters that are well suited for computations pertaining to the domain. Parameters such as size of the memory at different levels and bandwidth of the memory that is nearest to the compute resources is well experimented in the literature \cite{lac1}. Through pipelining of the processor and memory subsystem it is ensured that the processor is able to operate at the highest possible speed with lowest power penalty for the technology node \cite{pipeline1}\cite{pipeline2}. Several design space exploration techniques are developed to arrive at an optimum architectural parameters for optimal performance in the domain. These techniques are computer architecture simulator based techniques and allow tweaking of parameters such as memory size and memory bandwidth.

Basic Linear Algebra Subprograms (BLAS) and Linear Algebra Package (LAPACK) and/or their platform dependent variants are the basic building block for several high level software packages like Intel's DAAL, Spark's MLlib, Berkeley's CAFFE, UTK's PLASMA, and MAGMA packages \cite{caffe1}\cite{magma1}. Performance of BLAS and LAPACK eventually decides performance of these packages. Hence, it is important to have a high performance realization of these packages. Efficient realization of BLAS and LAPACK on different contemporary platforms has been ever researched topic \cite{mudge3}\cite{lac1}\cite{exp1}. All these efforts of efficient realizations are through software optimizations and efficient exploitation of memory hierarchy \cite{cgr1}\cite{Merc1}.   
Major reason for centralization of efforts toward software optimizations and efficient exploitation of memory hierarchy is mainly due to several architectural parameters that are not in the control of programmer \cite{exp3}. For example, the depth of the pipeline (pipeline stages) in the underlying platform \cite{fpu2}. In this paper, we present a theoretical framework that assists in establishing a relation between pipeline depth of different floating point operations with size and type of the workload. Major contributions in this paper are as follows:
\begin{itemize}
\item We present a comprehensive theoretical framework that allows us to predict processor performance based on pipeline depths of different floating point operations like multiplier, adder, square root, and divider for BLAS and LAPACK
\item Characterization of BLAS and LAPACK is presented where we try to determine several parameters to be fitted in our theoretical framework to arrive at optimum number of pipeline stages for floating point operations
\item Extensive simulations are carried out to arrive at an optimum pipeline depth of multiplier, adder, suquare root, and divider for BLAS and LAPACK in a Processing Element (PE). It is shown that our theoretical curves corroborate to our simulations. Finally with synthesis results it is shown that our PE outperforms recently presented custom linear algebra accelerator
\end{itemize}

We choose a scalar processor for our initial theoretical framework and then extend framework for superscalar processor. The paper is organized as follows:
In section \ref{sec:rw}, we discuss some of the works in the literature focusing on optimum pipeline depth of the processor. In section \ref{sec:tf}, we focus on theoretical framework and derive expression for optimum pipeline depth for several operations encountered in BLAS and LAPACK. Characterization of BLAS and LAPACK is presented in section \ref{sec:pe}. We present a Processing Element (PE) design in section \ref{sec:pe} for experimental setup that is to validate our theoretical framework and discuss results in section \ref{sec:res}. In section \ref{sec:con}, we conclude our work.

\section{Related Work}\label{sec:rw}
There is a significant theoretical and experimental work done in the recent past that establlishes relation between pipeline depth of a microprocessor and cache size \cite{pipeline1}\cite{pipeline2}.

In \cite{pipeline1}, authors have presented interesting work that focuses on improving processor performance by having deeper pipeline considering Intel Pentium $4$ as a baseline case. Relation between processor performance, pipeline depth, and cache size is established for several benchmarks. The paper presents simulator based experimental results. It is concluded that with 100\% increase in the performance in the Pentium $4$ like processors, performance improvement of 35-90\% can be attained. A major shortcoming of the work presented in \cite{pipeline1} is that the work presents interesting empirical results and does not establish succinct theory for predicting performance by varying pipeline depth and cache size.

In \cite{pipeline2}, authors have presented an analytical model that derives optimal pipeline depth as a function of power and performance for a superscalar processor. The model is validated using a cycle accurate simulator of a contemporary superscalar processor. Authors in \cite{pipeline2} build on the base case presented in \cite{pipeline6} where it is shown that for $s_i$ pipeline stages, if $t_i$ is the latch free time to complete the operation in pipe $i$, then in the scenario where all the pipe stages operate at same frequency, $\frac{t_i}{s_i} = \frac{t_j}{s_j}$, $\forall i,j$. If $c_i$ is latch overhead in $i^{th}$ pipeline stage than time per stage of pipe $i$ is $T_i  = \frac{t_i}{s_i} +  c_i$, $\forall i$. In case of absence of pipeline stalls, throughput of such a machine would be $G = \sum_{i=1}^k(\frac{1}{T_i})$, where $k$ is number of pipe stages in the pipeline. In \cite{pipeline2}, authors have extended this baseline model to incorporate pipeline stalls. The work presented in \cite{pipeline2} becomes one of the starting point for the work presented in this paper.   

In \cite{pipeline5}, authors have analyzed trade-off between greater throughput in deeper pipeline and penalty due to hazards in deeper pipeline. Sensitivity in Cycles-per-Instruction and cycle time are considered as parameters to arrive at optimum pipeline depth. It is shown that the total time can be modeled as a sum busy and non busy time of the pipeline considering pipeline hazards as a parameter. Simulation is performed for $35$ different types of workloads and it is clearly shown that the optimum pipeline depth varies between $13$ to $35$ for these workloads. Such a revelation gives us motivation to work further on a class of workloads for the workload specific (or domain specific) accelerator. The theoretical framework presented in \cite{pipeline5} forms foundation of our theoretical framework and the framework presented in \cite{pipeline5} is revisited in the prelude of section \ref{sec:tf}.

Theoretical framework presented in \cite{pipeline3} is continuation of the theoretical framework presented in \cite{pipeline5}. In \cite{pipeline3}, authors have optimized pipeline for power and performance considering $55$ workloads. The problem of optimum pipeline depth is well studied by considering parameters like dynamic power increase, clock gating, and leakage power in \cite{pipeline3}.

In \cite{lac4}, authors have presented several floating point unit architecture extensions to accelerate matrix factorizations. The work presented in \cite{lac4} is interesting and through several extension to the floating point unit architecture, significant performance improvement over baseline accelerator is achieved. The limitation of the work presented in \cite{lac4} is lack of theoretical framework that helps to decide the architectural parameters. The work presented in \cite{lac4} serves as a major benchmark for the work presented in this paper. 

In this paper, we have considered several theoretical and experimental framework as a motivation and/or baseline for our theoretical framework. We dwell on the idea of arriving at optimum pipeline depth for the domain customized accelerator. We perform analysis of the workload which is BLAS and LAPACK in this case and based on that we arrive at optimum pipeline depth of multiplier, adder, square root, and divider for the accelerator. Number of independent operations in the Directed Acyclic Graphs (DAGs) of the several routines BLAS and LAPACK are considered as parameters for floating point unit co-design for domain specific accelerator.

\section{Theoretical Framework}\label{sec:tf}
In the initial part of this section, we revisit theory presented in \cite{pipeline1}, \cite{pipeline2}, and \cite{pipeline5}. Latter we extend theory for domain customized architectures by considering workload characterization. The total time $T$ for the pipeline of the processor can be given by

\begin{align}\label{eqn:eq1}
	T = T_{BZ} + T_{NBZ}
\end{align}
where $T_{BZ}$, and $T_{NBZ}$ represent busy and non-busy time respectively.  Typically, $T_{BZ}$ is when pipeline is busy while $T_{NBZ}$ is when pipeline is stalled due one of the hazards. From \cite{pipeline5}, ratio of total time $T$ to the total number of instructions $N_I$ is given by

\begin{align}\label{eqn:eq8}
	\frac{T}{N_{I}} = (t_{o} + \frac{\gamma N_{H}t_{p}}{N_{I}}) + (\frac{t_{p}}{p}) + (\frac{\gamma N_{H}t_{o}p}{N_{I}})
\end{align}

In equation \ref{eqn:eq8}, $t_p$ is the total logic delay of the processor, $p$ is the number of pipeline stages in the design, $t_o$ is the latch overhead for the technology, $N_I$ is total number of instructions, $N_H$ is total number of pipeline hazards, and $\gamma =  \frac{1}{N_H}\sum^{N_H}\beta_h$ where $\beta_h$ is the fraction of the total pipeline delay encountered by each particular hazard.  

In equation \ref{eqn:eq8}, the first term is independent of pipeline depth; the second term varies inversely with $p$; and the last term varies linearly with $p$. To obtain minimum at particular value of $p$, equation \ref{eqn:eq8} can be differentiated and equated to $0$. That will give $p_{opt}$ 

\begin{align}\label{eqn:eq9}
	p^2_{opt} = \frac{N_{I}t_{p}}{\gamma N_{H}t_{o}}
\end{align}

Few observations about optimum pipeline depth can be made from equation \ref{eqn:eq9}. As $t_{o}$ which is latch overhead decreases with lowering node of technology, optimum pipeline depth increases. Lower the hazards in the workload the pipeline depth increases. As $\gamma$ which is fraction of the pipeline that hazards stall decreases, the optimum pipeline depth increases. 

\begin{figure}[!ht]
	\centering
	\includegraphics[scale = 0.22]{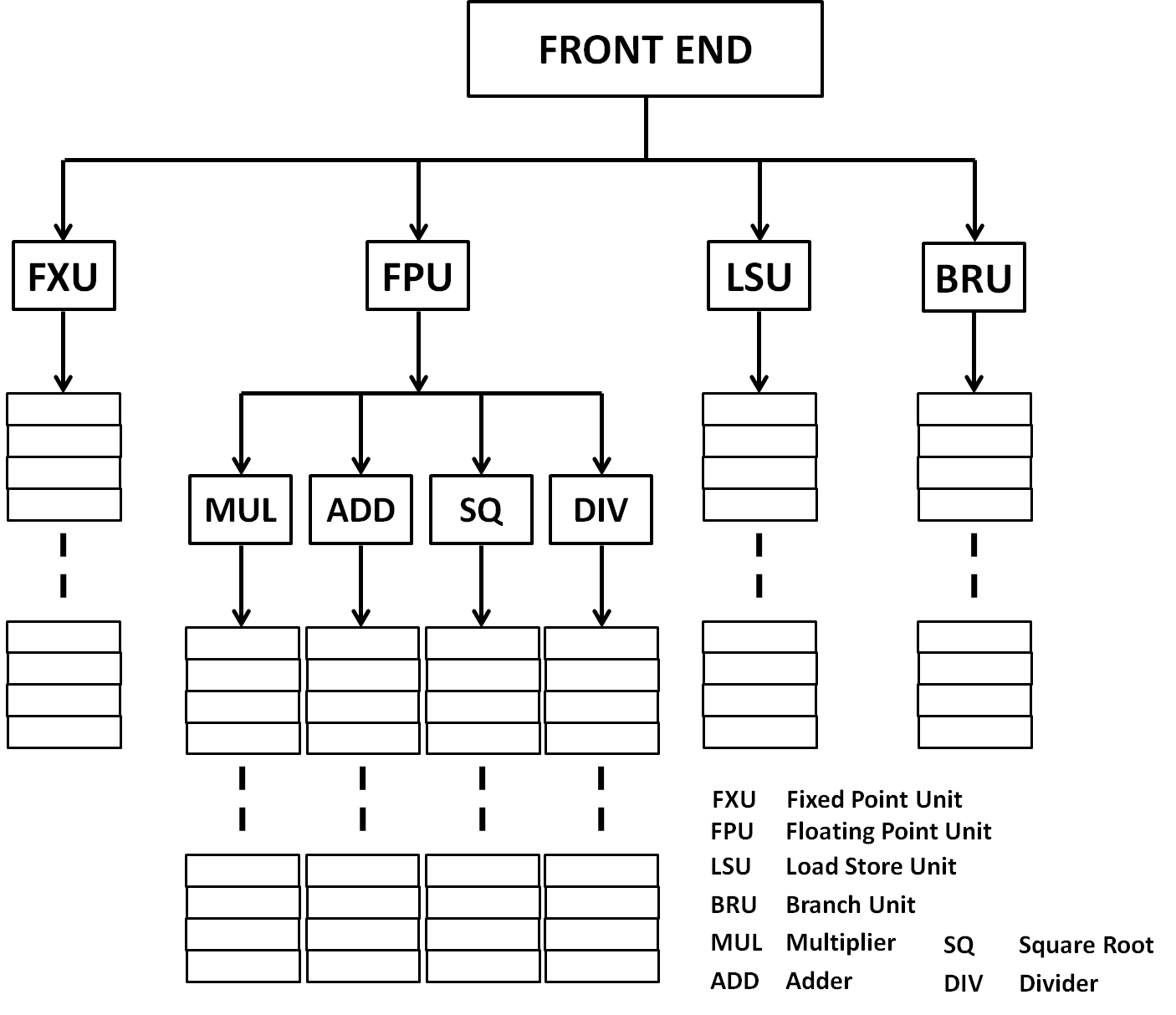}
	\caption{Pipeline Model}
	\label{fig:pipe1}
\end{figure}

We extend this theory for BLAS and LAPACK through workload characterization where we consider characteristics of the specific workload to arrive at an optimum pipeline depth of different operations in encountered in the workload. To extend theoretical frame work, we consider analytical pipeline model presented in \cite{pipeline2} that encompasses several pipes namely fixed point unit pipe, load-store pipe, and branch pipe. We extend the theoretical model presented in \cite{pipeline2} and incorporate a floating point pipe as shown in figure \ref{fig:pipe1}.

As shown in the figure \ref{fig:pipe1}, the model has four pipes: fixed point, floating point, load store, and branch. Since, in BLAS and LAPACK, the operations are floating point in nature and the operations encountered are multiply, addition, division, and square root, we further divide floating point unit pipeline into multiplier pipe, adder pipe, divide pipe, and square root pipelines. Our objective is to arrive at an optimum pipeline depth of these floating point hardware units. The types of arithmetic instructions encountered in BLAS and LAPACK can be given by a set $K = \{M,A,S,D\}$ where $M$, $A$, $S$, and $D$ are for multiplication, adder, square root and divider insturctions respectively. The total number of instructions in a routine of BLAS and/or LAPACK is given by 

\begin{align}
	N_{I} = \sum^K_i N_{iI} \text{  where } i\in K
\end{align}

Similarly, total number of hazards are given by 

\begin{align}
	N_{H} = \sum^K_i N_{iH} \text{  where } i\in K
\end{align}
To arrive at an optimum pipeline depth of the each individual pipes shown in the figure \ref{fig:pipe1}, we can replace $N_I$ and $N_H$ by corresponding pipe parameters. From equation \ref{eqn:eq8}, Time per Instruction (TPI) is given by 

\begin{align}
	TPI = \sum^K_i \frac{T_i}{N_{iI}} \text{  where } i\in K
\end{align}
where ${T_i} = (t_{o} + \frac{\gamma N_{iH}t_{p}}{N_{iI}}) + (\frac{t_{p}}{p}) + (\frac{\gamma N_{iH}t_{o}p}{N_{iI}}),\text{ }i\in K$. $T_M$, $T_A$, $T_D$, and $T_S$ are the total execution times for multiplier, adder, divider, and square root pipelines for an instruction stream. Parameter $t_o$ is technology dependent and not dependent on the type of the instruction. Equation \ref{eqn:eq9} can be modified as


\begin{align}
	\label{eqn:eqn_total}
	p^2_{opt_i} = \frac{N_{iI}t_{p_i}}{\gamma_i N_{iH}t_{o}} \text{ where } i \in K
\end{align}


In equation \ref{eqn:eqn_total}, $p_{M}$, $p_{A}$, $p_{D}$, and $p_{S}$ is the total number of pipeline stages in multiplier, adder, divider, and square root hardware unites respectively. Similarly, $\gamma_M$, $\gamma_A$, $\gamma_D$, and $\gamma_S$ are the total pipeline delay for each pipeline averaged over total number of hazards for each pipe. From \cite{pipeline5}, $\gamma = \frac{1}{N_H}\sum^{N_H}\beta_h$ where $\beta_h$ is fraction of total pipeline delay encountered by each particular hazard.  

In general, in absence of workload characterization, we can vary different parameters like $\gamma$, $N_I$, $N_H$, and $p$ in equation \ref{eqn:eq8} and comment on effect of different parameters on the time $T$.

\begin{figure}[!ht]
	\centering
	\includegraphics[scale = 0.22]{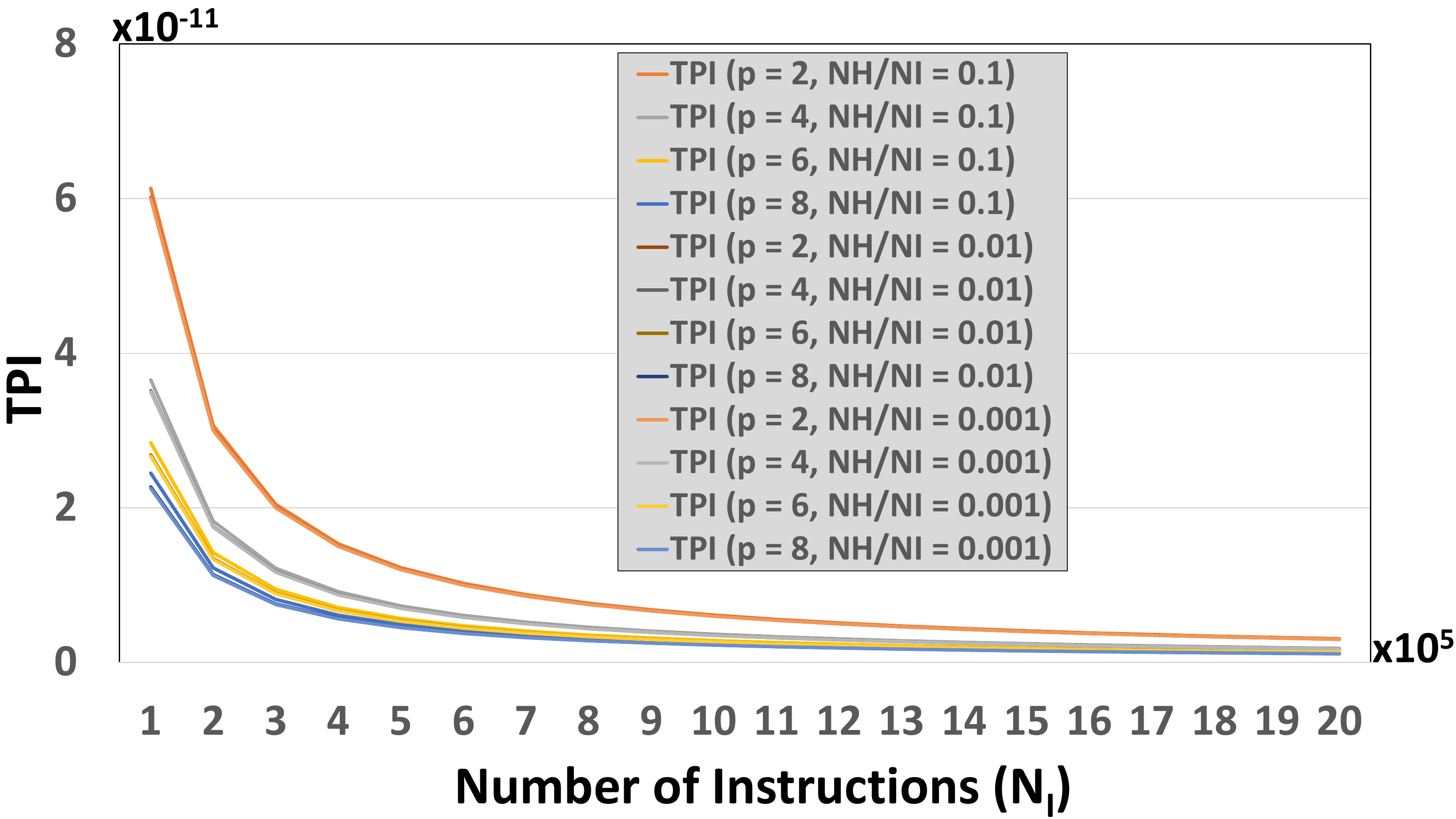}
	\caption{TPI for Different Sizes of Workload for $2,4,6,$ and $8$ (keeping $\frac{N_H}{N_I} = 0.1, 0.01,$ and $0.001$)}
	\label{fig:graph1}
\end{figure} 

In figure \ref{fig:graph1}, it can be observed that for a fixed number of pipeline stages $p$, as the problem size increases, the TPI saturates. For example, $p=2$ and $\frac{N_H}{N_I} = 0.1$, $0.01$, and $0.001$ then TPI saturates at instruction count of $10\times 10^5$ in the workload. This is mainly because smaller pipelines require large number of instruction to saturate and approach lower bound of TPI. It can also be observed in the figure \ref{fig:graph1} that for relatively larger pipelines (for $p=4$, $6$, and $8$) attained TPI progressively increases. This is mainly because of increased operating frequency of the pipeline stages. 

\begin{figure}[!ht]
	\centering
	\includegraphics[scale = 0.22]{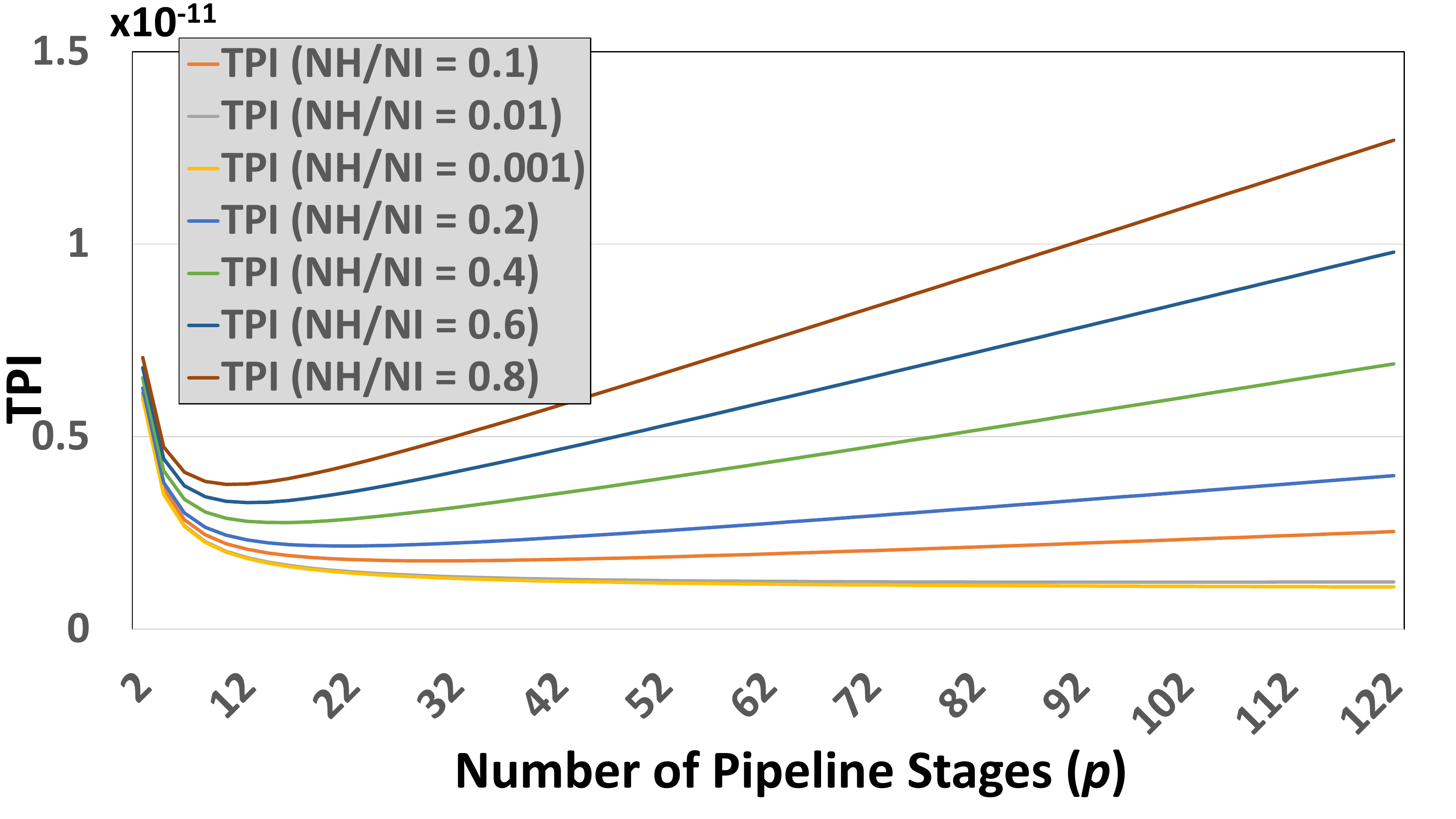}
	\caption{TPI for Different Pipeline Stages $p$ and Varying Workload (keeping $\frac{N_H}{N_I} = 0.1, 0.01, 0.001, 0.2, 0.4, 0.6,$ and $0.8$)}
	\label{fig:graph2}
\end{figure} 

Effect on TPI of varying pipeline depth for a particular workload with varying hazards is shown in figure \ref{fig:graph2}. It can be observed in the figure \ref{fig:graph2} that as we increase pipeline depth, TPI decreases and optimum is achieved. Beyond optimum, a linear increase in the TPI is observed. It can also be observed that the theoretical curve presented in \ref{fig:graph2} is fairly flat around optimum leaving considerable scope in choosing best design point for the optimum pipeline depth. 

\begin{figure}[!ht]
	\centering
	\includegraphics[scale = 0.22]{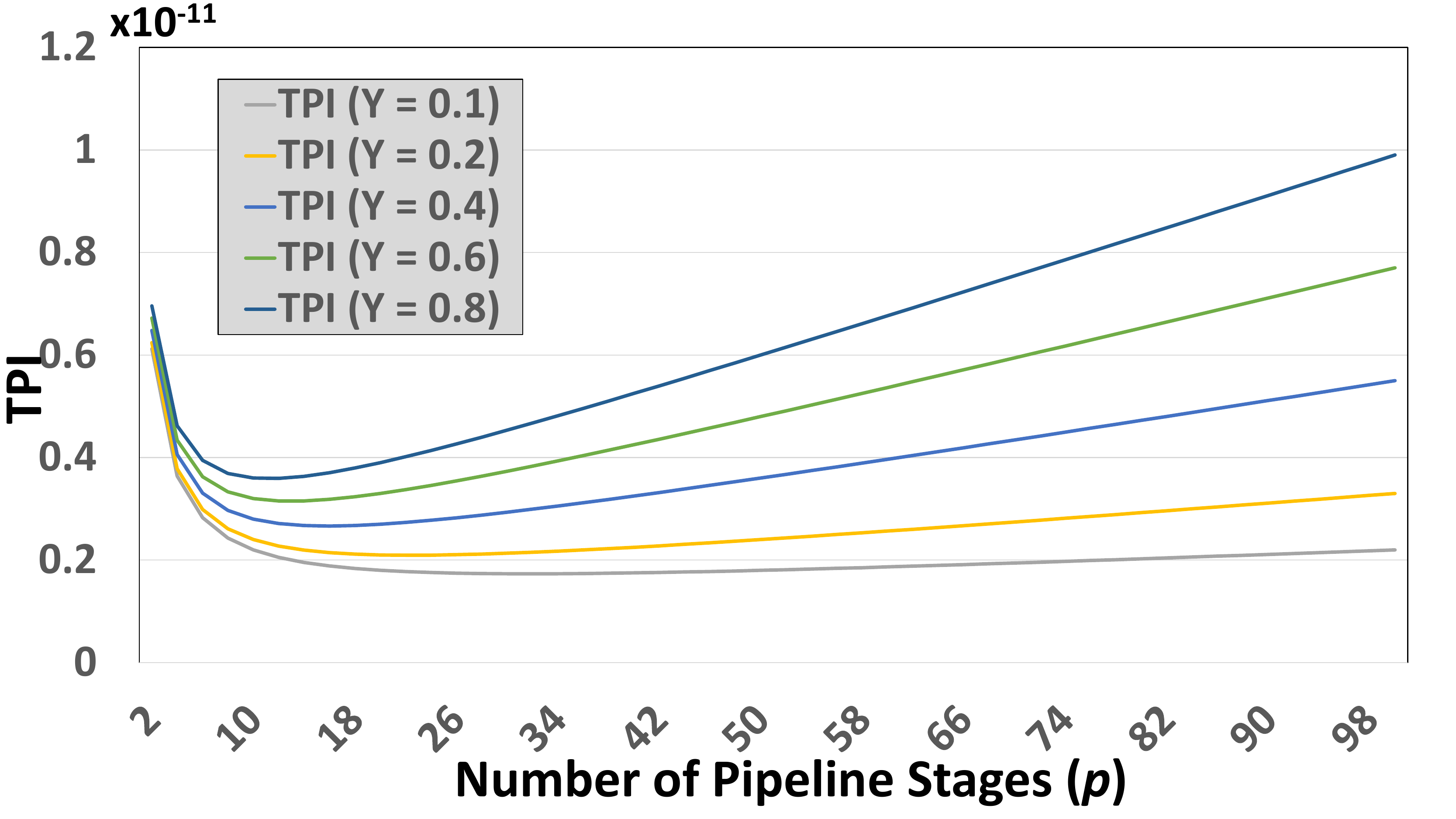}
	\caption{TPI for Different Pipeline Stages $p$ and $\gamma = 0.1, 0.2, 0.4, 0.6,$ and $0.8$}
	\label{fig:graph3}
\end{figure} 

Effect of varying $\gamma$ and pipeline stages $p$ on TPI is shown in figure \ref{fig:graph3}. It can be observed in the figure \ref{fig:graph3} that for a smaller values of $\gamma$, optimum achieved in the theoretical curve is around $4$ and as we increase value of $\gamma$, a deeper pipeline becomes optimum pipeline. From the figures \ref{fig:graph1}, \ref{fig:graph2}, and \ref{fig:graph3}, we can make following remarks:

\noindent{\bf Remark 1:}  Pipeline will saturate as we increase the size of the workload. Higher the ratio $\frac{N_H}{N_I}$, worse TPI is attained for small size of workloads.

\noindent{\bf Remark 2:} Higher the ratio $\frac{N_H}{N_I}$, shallow the optimum pipeline depth for the workload. It is better to have less number of pipeline stages if workload contains large number of hazards. For large number of hazards, if pipeline stages are higher than the optimum pipeline stages then the TPI attained deteriorates significantly as shown by red line (for $\frac{N_H}{N_I} = 0.8$) in the figure \ref{fig:graph2}

\noindent{\bf Remark 3:} Parameter $\gamma$ that solely depends on the total number of hazards $N_H$ and $\beta_h$ which is fraction of the total pipeline delay encountered by each particular hazard plays an important role in determination of optimum pipeline depth. For large value of $\gamma$, if the pipeline stages are more than $20$ and increased further, TPI deteriorates significantly as shown by blue line in the figure \ref{fig:graph3}. For small value of $\gamma$, even if the number of pipeline stages are increased beyond optimum number, the increase in TPI is observed minimal


Based on the observations from the theoretical curves in the figures \ref{fig:graph1}, \ref{fig:graph2}, and \ref{fig:graph3}, we can establish that it is important to characterize workloads of the domain of interest to arrive at an optimum pipeline depth of the different operations encountered in the computations pertaining to the domain. 

%
%
%
%

\section{BLAS and LAPACK Characterization}\label{sec:pe}
Based on remarks in section \ref{sec:tf} and theory presented in \cite{pipeline1}, and \cite{pipeline2}, we present detailed characterization of different routines in BLAS and LAPACK for determining several parameters that help us arriving at optimum pipeline depths of multiplier adder, square root and divider for these packages.

\subsection{Characterization of BLAS} \label{sec:4_1}
For characterization of BLAS, we consider $inner$ $product$ (Level-1 BLAS), $matrix-vector$ multiplication (Level-2 BLAS), and $general$ $matrix-matrix$ multiplication (Level-3 BLAS) as representative routines. These routines are known as $ddot$, $dgemv$, and $dgemm$ respectively where 'd' is for double precision \cite{laug}. 

For vectors $x = \begin{bmatrix} a_{1} & a_{2} & ... & a_{n} \end{bmatrix}$, and $y = \begin{bmatrix} b_{1} & b_{2} & ... & b_{n} \end{bmatrix}$, inner product is given by 

\begin{align}	
	c &= x^Ty \nonumber \\
	 &= \begin{bmatrix} a_{1} & a_{2} & a_{3} & a_{4} \end{bmatrix} \begin{bmatrix} b_{1} \\ b_{2}  \\ b_{3} \\ b_{4} \end{bmatrix} \text{for } n=4
\end{align}

\begin{figure}[!ht]
	\centering
	\includegraphics[scale = 0.32]{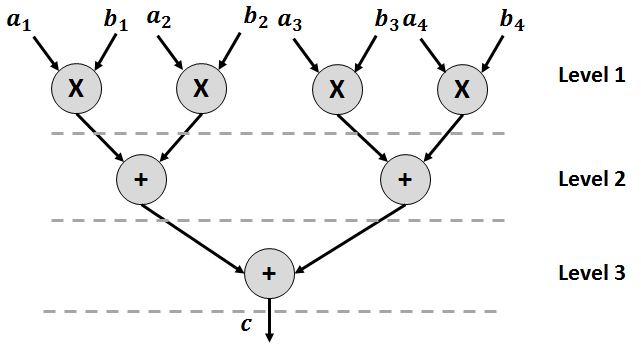}
	\caption{4-element Vector Inner Product}
	\label{fig:dot4_ex}
\end{figure}

Directed Acyclic Graph (DAG) for $n=4$ is shown in figure \ref{fig:dot4_ex}. It can be observed in the figure \ref{fig:dot4_ex} that all the multiplications in the $inner$ $product$ of $4-$element vector can be performed in parallel. In general for $n-$element vector there are $n$ multiplications and all the multiplications can be executed in parallel. There are $n-1$ additions in the $inner$ $product$, and there is a dependency from the output of the multiplier for the first level of the addition as shown in the figure \ref{fig:dot4_ex} and there are dependencies in the addition for each next level from the additions in the previous level. Considering, only dependency hazards, there will be no hazards in the multiplier pipeline. Associated parameters with multiplier, and adder pipelines shown in the figure \ref{fig:pe} will be as follows:

\begin{align}	
&	N_I = N_{IM}+N_{IA} = n + n -1 = 2n-1 \nonumber \\
&	N_{HM} = 0 \text{  (considering only dependency hazards)} \nonumber \\
&	\gamma_M = \infty \nonumber 
\end{align}

\begin{figure}[!ht]
	\centering
	\includegraphics[scale = 0.22]{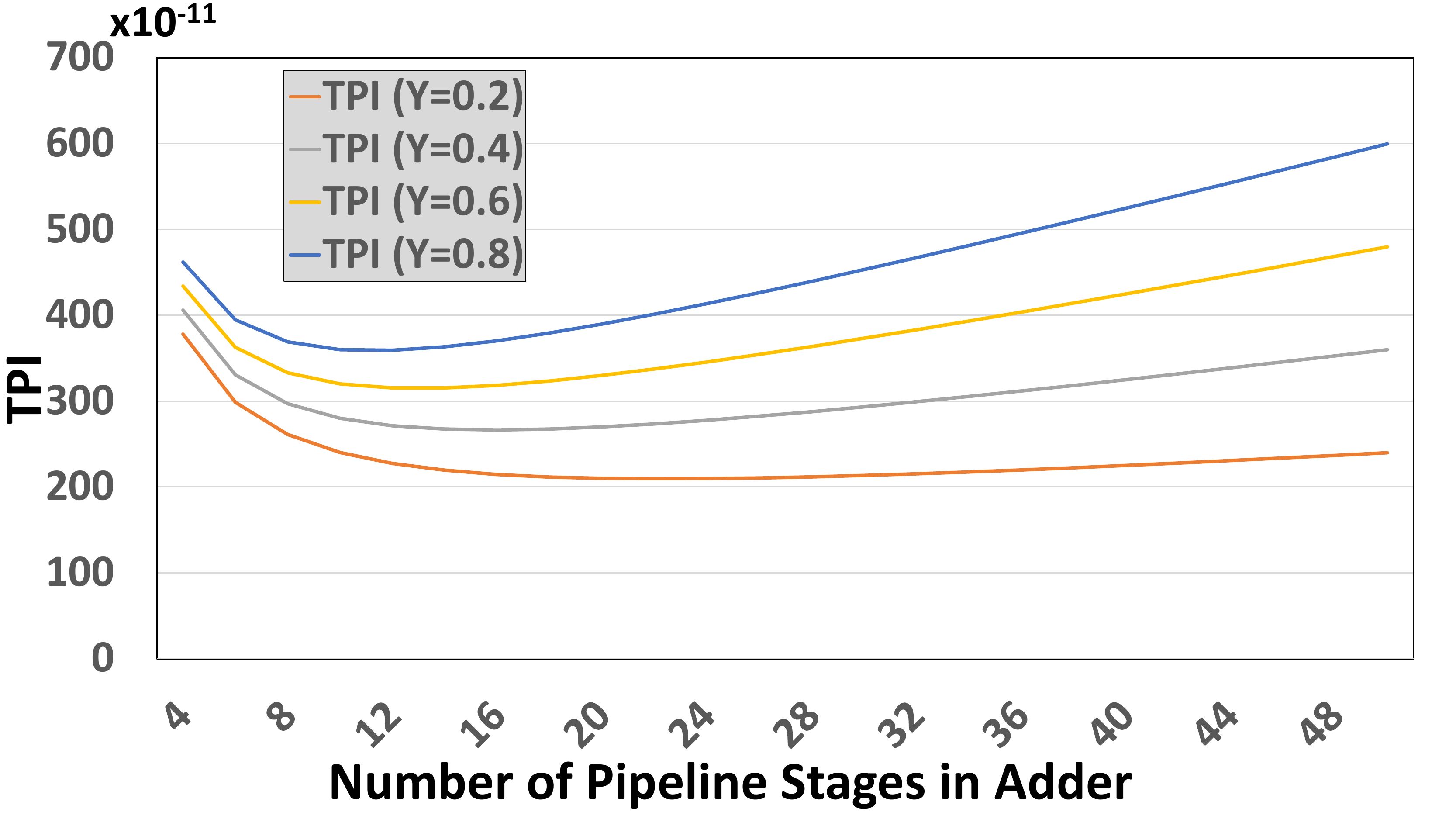}
	\caption{TPI for Different Pipeline Stages $p$ in Adder and $\gamma = 0.2, 0.4, 0.6,$ and $0.8$ for $1000-$ Element Vector Inner Product}
	\label{fig:graph4}
\end{figure} 

\begin{figure}[!ht]
	\centering
	\includegraphics[scale = 0.22]{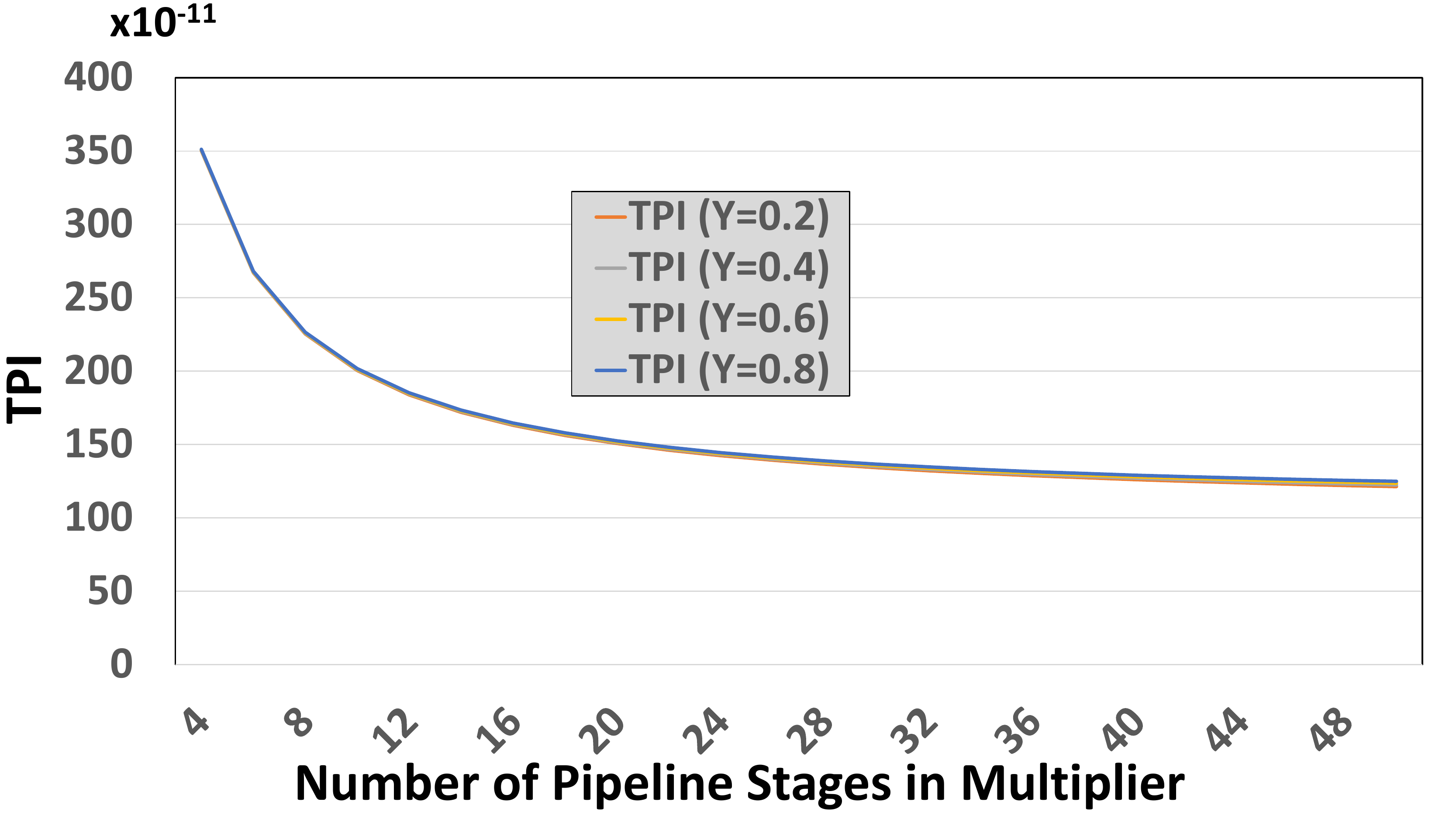}
	\caption{TPI for Different Pipeline Stages $p$ in Adder and $\gamma = 0.2, 0.4, 0.6,$ and $0.8$ for $1000-$ Element Vector Inner Product}
	\label{fig:graph5}
\end{figure} 

\begin{figure}[!ht]
	\centering
	\includegraphics[scale = 0.22]{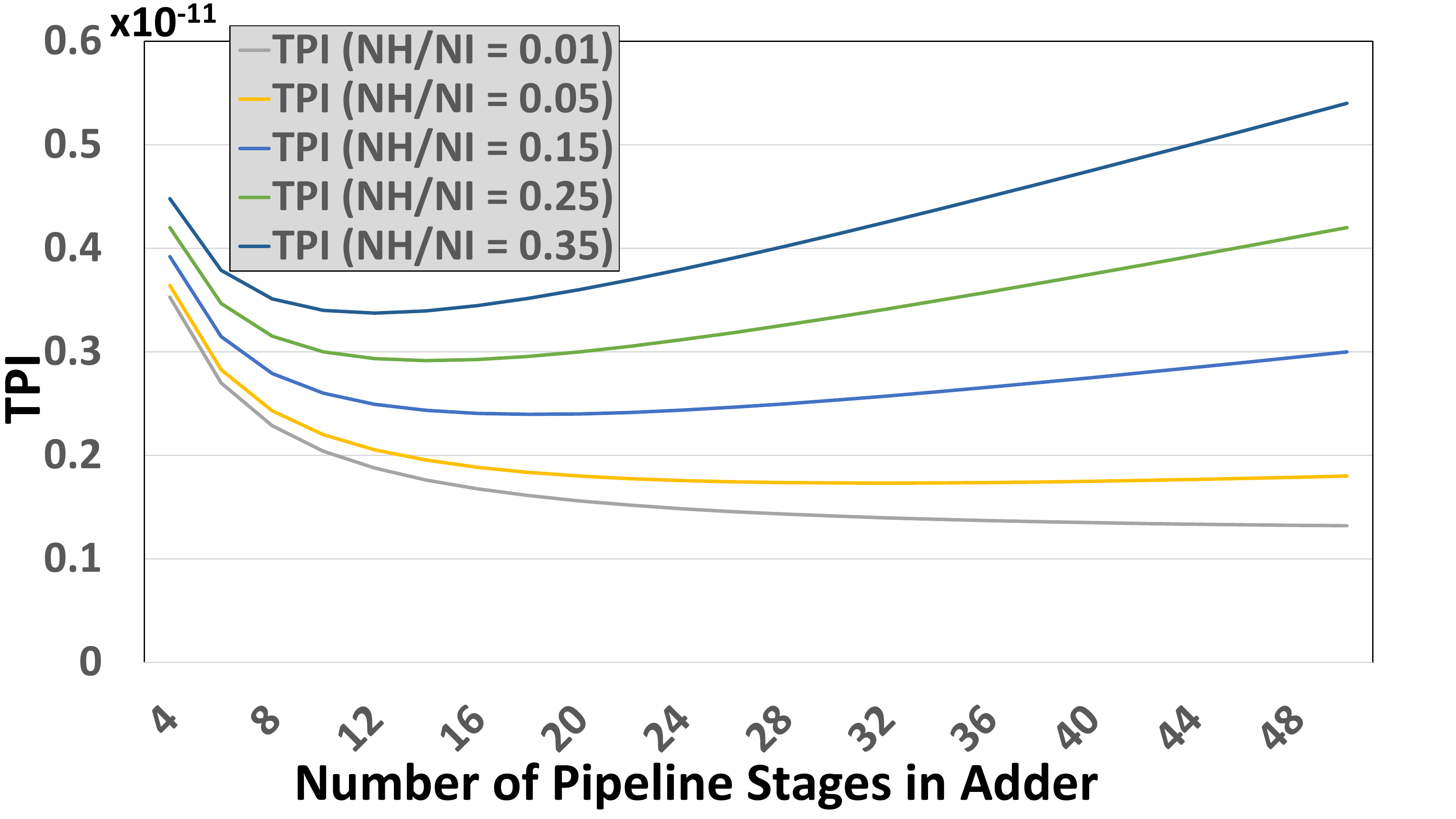}
	\caption{TPI for Different Pipeline Stages $p$ in Adder and $\frac{N_H}{N_I} = 0.01, 0.05, 0.15, 0.25,$ and $0.35$}
	\label{fig:graph6}
\end{figure} 

Determining $\gamma_A$ is difficult as mentioned in \cite{pipeline5}. Hence, we have to determine value of $\gamma_A$ through a theoretical curve shown in figure \ref{fig:graph4}. It can be observed that for large value of $\gamma_A$, a sharp rise in TPI is observed. For small value of $\gamma_A$, the curve becomes almost flat. Near optimum value in the curve, it is considerably flat allowing designer multiple choices for the number of pipeline stages. For figure \ref{fig:graph4}, we have considered $\frac{N_H}{N_I} = 0.1$. Decreasing $\frac{N_H}{N_I}$ further gives a flat theoretical curve as observed in the figure \ref{fig:graph2}. For multiplier, theoretical curve for TPI becomes a flat horizontal line as we increase the pipeline depth. This is mainly due to absence of dependency hazards in the multiplication. 

For $matrix-vector$, and $matrix-matrix$ multiplication, 
\begin{align}	
	y = Ax \\
	C = AB
\end{align}
where $y$ and $x$ are vectors, and $A$, $B$, and $C$ are matrices. Since, $matrix-vector$ multiplication, and $matrix-matrix$ multiplication can be viewed as a series of calls of $inner$ $products$, the optimum number of pipeline stages for these routines for adder and multiplier are expected to be the same as what we achieved for $inner$ $product$. It is well established that , in practical implementations of $matrix-vector$ multiplication (DGEMV in BLAS) and $matrix-matrix$ multiplication (DGEMM in BLAS), due to compiler optimizations the dependency hazards reduce \cite{nick1}. This reduction in the hazards will lead to increase in the $\gamma_A$ and decrease in the ratio $\frac{N_H}{N_I}$. In figure \ref{fig:graph6}, TPI for different ratio of $\frac{N_H}{N_I}$ is shown. It can be observed in the figure \ref{fig:graph6} that as the ratio $\frac{N_H}{N_I}$ increases, the growth in TPI is sharper.

\subsection{Characterization of LAPACK}

For LAPACK, we consider two most popular factorization routines namely DGEQRF (QR factorization), and DGETRF (LU factorization with partial pivoting) for characterization. 

\begin{figure}[!ht]
	\centering
	\includegraphics[scale = 0.30]{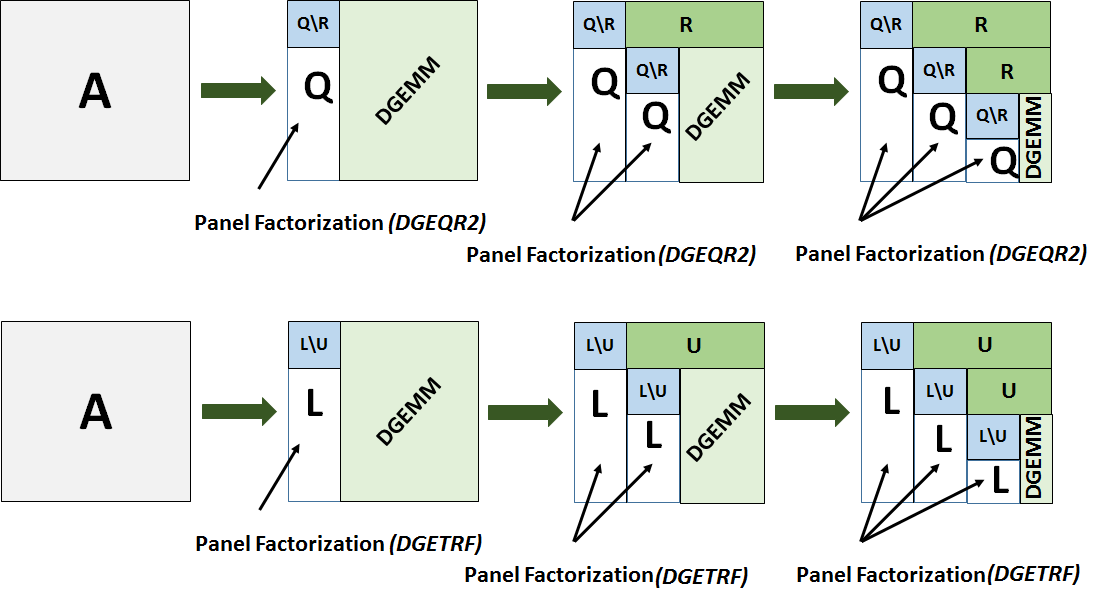}
	\caption{QR and LU Factorizations}
	\label{fig:lu_qr}
\end{figure} 

\begin{figure}[!ht]
	\centering
	\includegraphics[scale = 0.22]{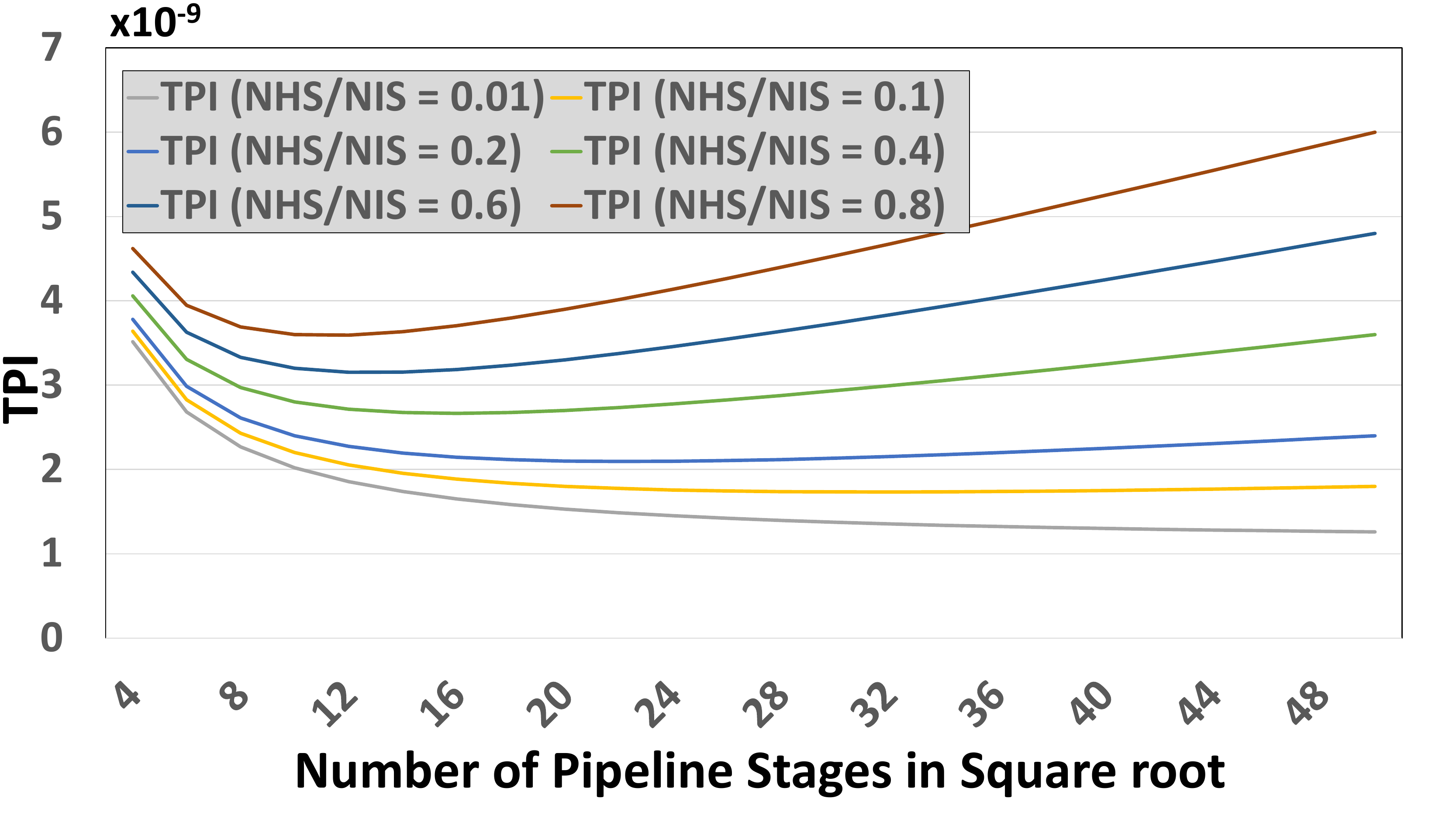}
	\caption{TPI for Different Pipeline Stages $p$ in Square root and $\frac{N_{HS}}{N_{IS}} = 0.01, 0.1, 0.2, 0.4, 0.6,$ and $0.8$}
	\label{fig:graph7}
\end{figure} 
 
For these factorizations, it can be observed in figure \ref{fig:lu_qr} that the $matrix-matrix$ operations (DGEMM) are dominant, and hence the optimum number of pipeline stages for multiplier and adder would remain same as derived in section \ref{sec:4_1}. It is important to arrive at an optimum pipeline depth of divider and square root shown in the figure \ref{fig:pe} through characterization. 

In QR factorization, division and square root operations are required in panel factorization and the order of division and square root operations is $O(n^2)$ while the total operations in the factorization are $O(n^3)$. There is always dependency in the square root operation that stalls the program execution. The ratios $\frac{N_{HD}}{N_{ID}}$, and $\frac{N_{HS}}{N_{IS}}$ are observed to be high in QR factorization. With varying pipeline and varying number of hazards in the square root pipeline $N_{HS}$, the theoretical curve is shown in figure \ref{fig:graph7}. For optimum number of stages in the divider, we expect trend that is similar to shown in the figure \ref{fig:graph7} since the number of dependency hazards in square root and divider are expected to be same in QR factorization. 

In LU factorization there are multiplications, additions, and divisions. Since the occurrence of division instruction in the program is similar to the square root/divider in the QR factorization, we expect similar trend for optimum pipeline stages for divider as shown in the figure \ref{fig:graph7}.

\section{Experimental Setup and Results}\label{sec:res}
\begin{figure}[!ht]
	\centering
	\includegraphics[scale = 0.27]{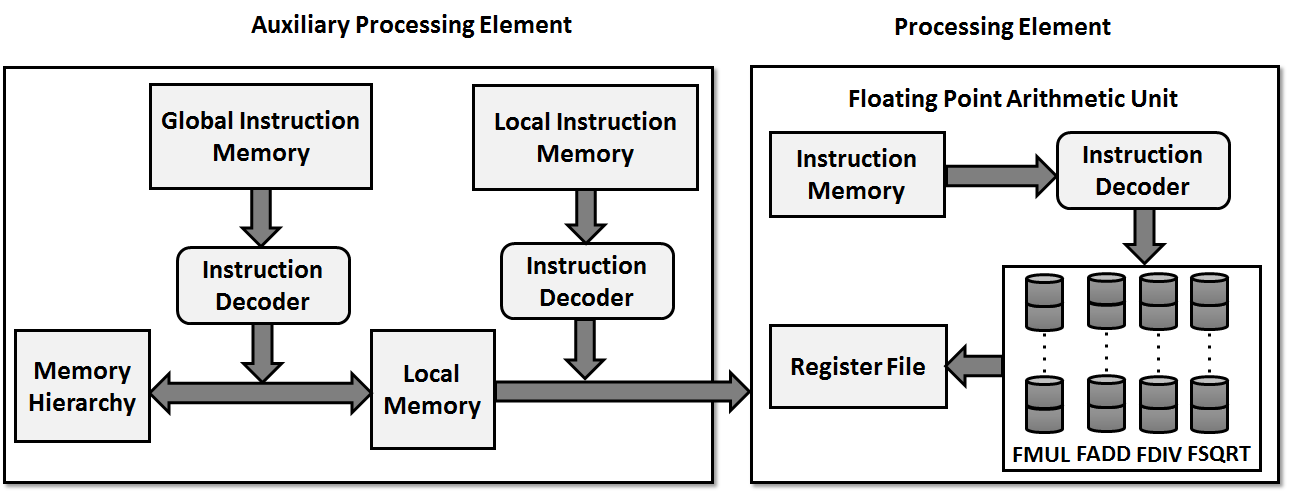}
	\caption{Experimental Setup}
	\label{fig:pe}
\end{figure} 

\begin{figure}[!ht]
	\centering
	\includegraphics[scale = 0.22]{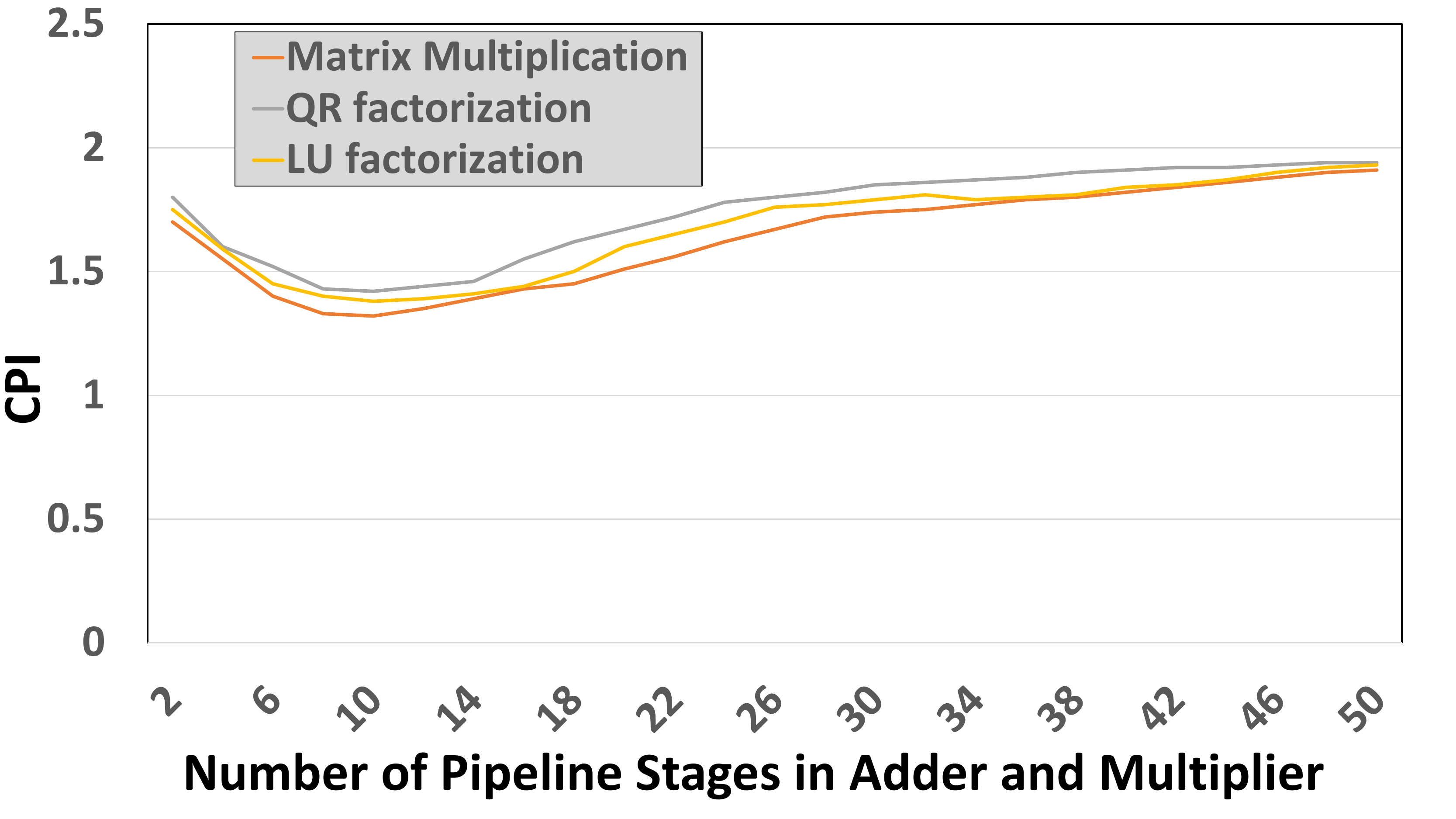}
	\caption{CPI in Matrix Multiplication, QR factorization, and LU factorization with Varying Number of Pipeline Stages in Adder and Multiplier for a Matrix of Size $100\times 100$}
	\label{fig:graph8}
\end{figure}

\begin{figure}[!ht]
	\centering
	\includegraphics[scale = 0.22]{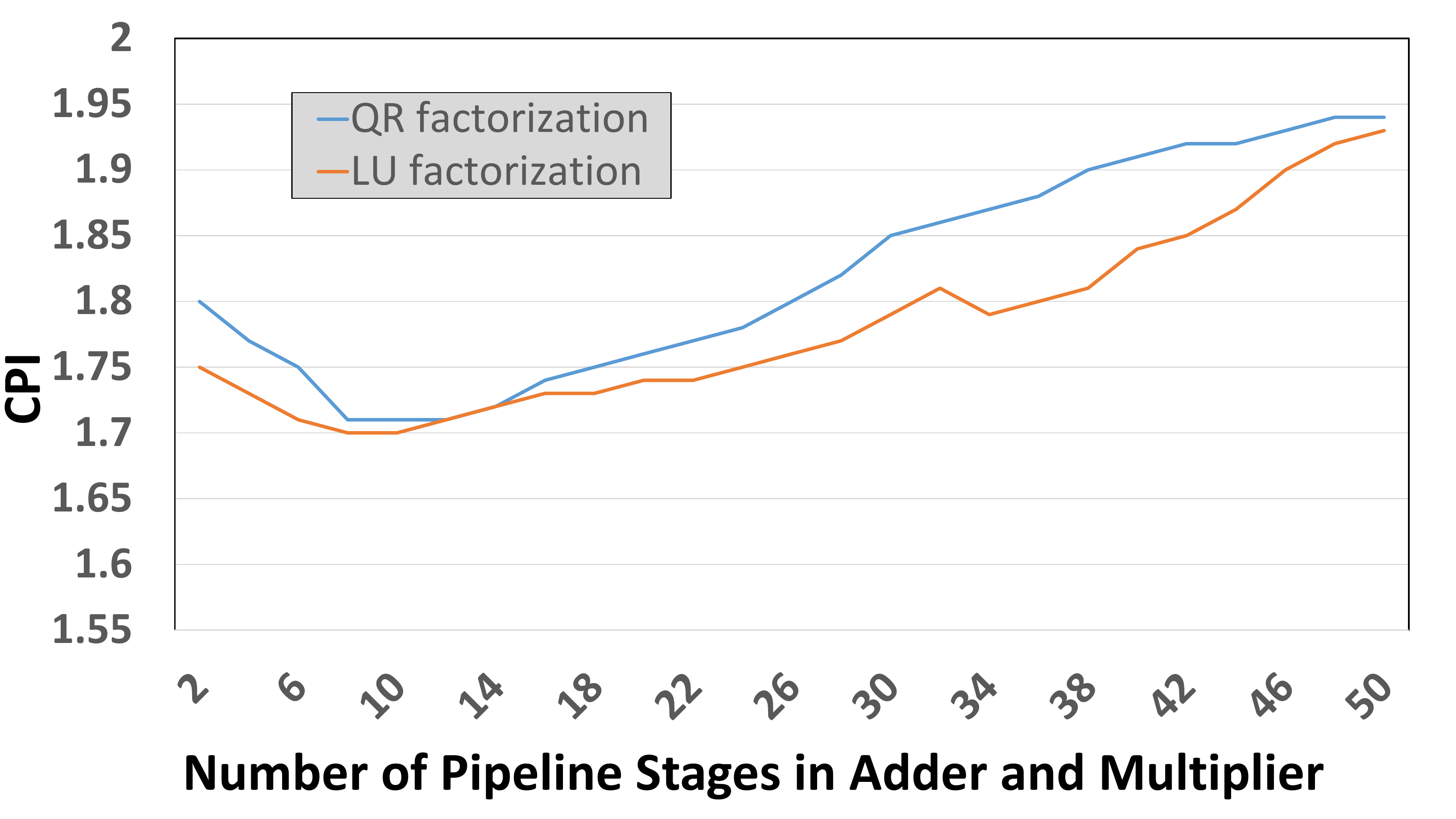}
	\caption{CPI in QR factorization, and LU factorization with Varying Number of Pipeline Stages in Square root and Divider for a Matrix of Size $100\times 100$}
	\label{fig:graph9}
\end{figure}

Experimental setup is shown in figure \ref{fig:pe}. As shown in the figure \ref{fig:pe}, we have a Processing Element (PE) and an Auxiliary Processing Element (APE) where PE has compute resources and load/store operations are handled by APE. PE consists of different floating point operations like multiplier, adder, square root, and divider, a small register file, an instruction memory, and an instruction decoder, and APE consists of two small instruction memories, corresponding instruction decoders, and a Local Memory (LM). The operation of PE and APE can be described as follows: \\

\noindent{\bf Step 1:} Load data from upper level of memory to LM in APE \\
\noindent{\bf Step 2:} Load data from LM to Register File \\
\noindent{\bf Step 3:} Perform computations in the PE and store results back in the Register File \\
\noindent{\bf Step 4:} Store results back from Register File to LM \\
\noindent{\bf Step 5:} Store final result to the upper level of memory \\


\begin{table}[!ht]
\begin{center}
\caption{Comparison between LAP-PE and PE at Different Frequencies with 16KBytes of dual-ported SRAM with double precision floating point arithmetic}
\begin{tabular}{ |p{1.8cm}|p{0.8cm}|p{0.8cm}|p{0.9cm}|p{0.8cm}|p{0.8cm}| } 
 \hline
  Architecture & Speed (GHz) & Area ($mm^2$) & Memory (mW) & FMAC (mW) & PE (mW) \\  \hline \hline
LAP-PE & 1.81 & 0.181 & 13.25 & 105.5 & 118.7 \\ \hline
LAP-PE & 0.95 & 0.174 & 6.95 & 31.0 & 38.0 \\ \hline
LAP-PE & 0.33 & 0.167 & 2.41 & 6.0 & 8.4 \\ \hline
LAP-PE & 0.20 & 0.169 & 1.46 & 3.4 & 4.8 \\ \hline
PE & 1.81 & 0.301 & 26.50 & 422 & 448.5 \\ \hline
PE & 0.95 & 0.28 & 13.90 & 124 & 137.9 \\ \hline
PE & 0.33 & 0.273 & 4.82 & 24 & 28.82 \\ \hline
PE & 0.20 & 0.275 & 2.92 & 13.6 & 16.5 \\ \hline    
\end{tabular}
\label{tab:res_pe}
\end{center}
\end{table}

\begin{table*}[!ht]
\begin{center}
\caption{Comparison of LAP-PE and PE}
\begin{tabular}{ |p{1.6cm}|p{2.8cm}|p{1.8cm}|p{1.8cm}|p{1.8cm}|} 
 \hline
  Speed & LAP-PE (GFlops/$mm^2$) & LAP-PE (GFlops/W) & PE (GFlops/$mm^2$) & PE (GFlops/W) \\  \hline \hline
1.81 & 19.92 & 29.7 & 42.09 & 28.24  \\ \hline
0.95 & 10.92 & 46.4 & 23.75 & 48.54  \\ \hline
0.33 & 3.95 & 57.8 & 8.46 & 82.5  \\ \hline
0.2 & 2.37 & 51.1 & 5.09 & 84.84  \\ \hline
\end{tabular}
\label{tab:res_pe1}
\end{center}
\end{table*}

As shown in the figure \ref{fig:pe}, the pipeline depths of the floating point arithmetic units are kept variable. Usually, it is not possible vary pipeline stages of the floating point unit in RTL. For simulation purpose, we use Bluespec System Verilog (BSV) that lets us incorporation of C program along with RTL registers to mimic the different number of pipeline stages for different floating point operations. The simulation environment also becomes non-synthesizable due to presence of floating point operation written in C. For simulation results, we report Cycles-per-Instructions (CPI) for varying number of pipeline stages for adder and multiplier for $matrix-matrix$ multiplication, QR factorization, and LU factorization as shown in figure \ref{fig:graph8}. It can be observed in the figure \ref{fig:graph8} that our simulation results corroborate to our theoretical curve observed in section \ref{sec:tf}.

For synthesis, we use enhanced version of PE where we attach 4 multipliers and 3 adders in a reconfigurable way to enhance the performance of the PE. Table \ref{tab:res_pe} presents comparison between LAP-PE and PE. It can be observed from the table \ref{tab:res_pe} that PE has more area and consumes more power. This is mainly because of SRAM and DOT4 instruction. If we take GFlops/$mm^2$ and GFlops/W as a performance measure as shown in table \ref{tab:res_pe1} then at 1.81 GHz, LAP-PE attains 19.92 GFlops/$mm^2$ while PE attains 42.09 GFlops/$mm^2$. Similarly, at 0.20 GHz, LAP-PE attains 2.37 GFlops/$mm^2$ while PE attains 5.09 GFlops/$mm^2$. 

Similarly, at 1.81 GHz, LAP-PE attains 29.7 GFlops/W while PE attains 28.281 GFlops/W. At 0.95 GHz, LAP-PE attains 46.4 GFlops/W while in PE it is 48.54 GFlops/W. At 0.2 GHz, LAP-PE achieves 51.1 GFlops/W while PE achieves 84.84 GFlops/W. 

It can be concluded from above observations that PE performs better than LAP-PE at lower frequencies. This is mainly because lower power consumed by double precision floating point operations at low frequencies.

\section{Conclusion}\label{sec:con}
We presented theoretical framework to arrive at an optimum number of pipeline stages for adder, multiplier, square root, and divider for BLAS and LAPACK. We presented characterization of BLAS and LAPACK to estimate parameters. The estimated parameters were used to arrive at theoretical curves. We also presented a PE that has extensible pipelines in the simulation environment. Through simulations, we show that our theoretical results corroborates to our simulation results. We synthesize PE with RTL of floating point unit and show better performance than the most recent custom realization of BLAS and LAPACK. Through our theoretical framework and experimental studies, it was shown that for domain specific platforms, it is possible and advisable to first derive an optimum pipeline depth theoretically for better performance of the platform. The theoretical framework presented can be extended with more precise determination of parameters like $\gamma$ and $N_H$. Near accurate determination of these parameters would result in better estimation of the optimum number of pipeline stages in domain specific platforms. 

\bibliographystyle{IEEEtran}
\bibliography{IEEEabrv,ref}

\begin{thebibliography}{10}
\providecommand{\url}[1]{#1}
\csname url@samestyle\endcsname
\providecommand{\newblock}{\relax}
\providecommand{\bibinfo}[2]{#2}
\providecommand{\BIBentrySTDinterwordspacing}{\spaceskip=0pt\relax}
\providecommand{\BIBentryALTinterwordstretchfactor}{4}
\providecommand{\BIBentryALTinterwordspacing}{\spaceskip=\fontdimen2\font plus
\BIBentryALTinterwordstretchfactor\fontdimen3\font minus
  \fontdimen4\font\relax}
\providecommand{\BIBforeignlanguage}[2]{{%
\expandafter\ifx\csname l@#1\endcsname\relax
\typeout{** WARNING: IEEEtran.bst: No hyphenation pattern has been}%
\typeout{** loaded for the language `#1'. Using the pattern for}%
\typeout{** the default language instead.}%
\else
\language=\csname l@#1\endcsname
\fi
#2}}
\providecommand{\BIBdecl}{\relax}
\BIBdecl

\bibitem{mudge3}
\BIBentryALTinterwordspacing
T.~N. Mudge, ``The specialization trend in computer hardware: techincal
  perspective,'' \emph{Commun. {ACM}}, vol.~58, no.~4, p.~84, 2015. [Online].
  Available: \url{http://doi.acm.org/10.1145/2735839}
\BIBentrySTDinterwordspacing

\bibitem{lac1}
A.~Pedram, S.~Z. Gilani, N.~S. Kim, R.~A. van~de Geijn, M.~J. Schulte, and
  A.~Gerstlauer, ``A linear algebra core design for efficient level-3 blas,''
  in \emph{ASAP}, 2012, pp. 149--152.

\bibitem{baluni1}
A.~Baluni, F.~Merchant, S.~K. Nandy, and S.~Balakrishnan, ``A fully pipelined
  modular multiple precision floating point multiplier with vector support,''
  in \emph{2011 International Symposium on Electronic System Design}, Dec 2011,
  pp. 45--50.

\bibitem{hyper1}
\BIBentryALTinterwordspacing
S.~Das, K.~T. Madhu, M.~Krishna, N.~Sivanandan, F.~Merchant, S.~Natarajan,
  I.~Biswas, A.~Pulli, S.~K. Nandy, and R.~Narayan, ``A framework for
  post-silicon realization of arbitrary instruction extensions on
  reconfigurable data-paths,'' \emph{Journal of Systems Architecture - Embedded
  Systems Design}, vol.~60, no.~7, pp. 592--614, 2014. [Online]. Available:
  \url{http://dx.doi.org/10.1016/j.sysarc.2014.06.002}
\BIBentrySTDinterwordspacing

\bibitem{lac3}
\BIBentryALTinterwordspacing
A.~Pedram, A.~Gerstlauer, and R.~A. van~de Geijn, ``Algorithm, architecture,
  and floating-point unit codesign of a matrix factorization accelerator,''
  \emph{{IEEE} Trans. Computers}, vol.~63, no.~8, pp. 1854--1867, 2014.
  [Online]. Available: \url{http://dx.doi.org/10.1109/TC.2014.2315627}
\BIBentrySTDinterwordspacing

\bibitem{Gregg2}
\BIBentryALTinterwordspacing
M.~H. Ionica and D.~Gregg, ``The movidius myriad architecture's potential for
  scientific computing,'' \emph{{IEEE} Micro}, vol.~35, no.~1, pp. 6--14, 2015.
  [Online]. Available: \url{http://dx.doi.org/10.1109/MM.2015.4}
\BIBentrySTDinterwordspacing

\bibitem{cgr2}
Z.~E. R{\'a}kossy, F.~Merchant, A.~A. Aponte, S.~K. Nandy, and
  A.~Chattopadhyay, ``Efficient and scalable cgra-based implementation of
  column-wise givens rotation,'' in \emph{ASAP}, 2014, pp. 188--189.

\bibitem{cgr3}
\BIBentryALTinterwordspacing
Z.~E. R{\'{a}}kossy, F.~Merchant, A.~A. Aponte, S.~K. Nandy, and
  A.~Chattopadhyay, ``Scalable and energy-efficient reconfigurable accelerator
  for column-wise givens rotation,'' in \emph{22nd International Conference on
  Very Large Scale Integration, VLSI-SoC, Playa del Carmen, Mexico, October
  6-8, 2014}, 2014, pp. 1--6. [Online]. Available:
  \url{http://dx.doi.org/10.1109/VLSI-SoC.2014.7004166}
\BIBentrySTDinterwordspacing

\bibitem{pipeline1}
\BIBentryALTinterwordspacing
E.~Sprangle and D.~Carmean, ``Increasing processor performance by implementing
  deeper pipelines,'' in \emph{Proceedings of the 29th Annual International
  Symposium on Computer Architecture}, ser. ISCA '02.\hskip 1em plus 0.5em
  minus 0.4em\relax Washington, DC, USA: IEEE Computer Society, 2002, pp.
  25--34. [Online]. Available:
  \url{http://dl.acm.org/citation.cfm?id=545215.545219}
\BIBentrySTDinterwordspacing

\bibitem{pipeline2}
\BIBentryALTinterwordspacing
V.~Srinivasan, D.~Brooks, M.~Gschwind, P.~Bose, V.~Zyuban, P.~N. Strenski, and
  P.~G. Emma, ``Optimizing pipelines for power and performance,'' in
  \emph{Proceedings of the 35th Annual ACM/IEEE International Symposium on
  Microarchitecture}, ser. MICRO 35.\hskip 1em plus 0.5em minus 0.4em\relax Los
  Alamitos, CA, USA: IEEE Computer Society Press, 2002, pp. 333--344. [Online].
  Available: \url{http://dl.acm.org/citation.cfm?id=774861.774897}
\BIBentrySTDinterwordspacing

\bibitem{caffe1}
Y.~Jia, E.~Shelhamer, J.~Donahue, S.~Karayev, J.~Long, R.~Girshick,
  S.~Guadarrama, and T.~Darrell, ``Caffe: Convolutional architecture for fast
  feature embedding,'' \emph{arXiv preprint arXiv:1408.5093}, 2014.

\bibitem{magma1}
B.~J. Smith, ``R package magma: Matrix algebra on gpu and multicore
  architectures, version 0.2.2,'' September 3, 2010, [On-line]
  http://cran.r-project.org/package=magma.

\bibitem{exp1}
\BIBentryALTinterwordspacing
M.~Mahadurkar, F.~Merchant, A.~Maity, K.~Vatwani, I.~Munje, N.~Gopalan, S.~K.
  Nandy, and R.~Narayan, ``Co-exploration of {NLA} kernels and specification of
  compute elements in distributed memory cgras,'' in \emph{XIVth International
  Conference on Embedded Computer Systems: Architectures, Modeling, and
  Simulation, {SAMOS} 2014, Agios Konstantinos, Samos, Greece, July 14-17,
  2014}, 2014, pp. 225--232. [Online]. Available:
  \url{http://dx.doi.org/10.1109/SAMOS.2014.6893215}
\BIBentrySTDinterwordspacing

\bibitem{cgr1}
\BIBentryALTinterwordspacing
F.~Merchant, A.~Chattopadhyay, G.~Garga, S.~K. Nandy, R.~Narayan, and
  N.~Gopalan, ``Efficient {QR} decomposition using low complexity column-wise
  givens rotation {(CGR)},'' in \emph{2014 27th International Conference on
  {VLSI} Design and 2014 13th International Conference on Embedded Systems,
  Mumbai, India, January 5-9, 2014}, 2014, pp. 258--263. [Online]. Available:
  \url{http://dx.doi.org/10.1109/VLSID.2014.51}
\BIBentrySTDinterwordspacing

\bibitem{Merc1}
F.~Merchant, A.~Maity, M.~Mahadurkar, K.~Vatwani, I.~Munje, M.~Krishna,
  S.~Nalesh, N.~Gopalan, S.~Raha, S.~Nandy, and R.~Narayan,
  ``Micro-architectural enhancements in distributed memory cgras for lu and qr
  factorizations,'' in \emph{VLSI Design (VLSID), 2015 28th International
  Conference on}, Jan 2015, pp. 153--158.

\bibitem{exp3}
F.~Merchant, T.~Vatwani, A.~Chattopadhyay, S.~Raha, S.~K. Nandy, and
  R.~Narayan, ``Achieving efficient qr factorization by algorithm-architecture
  co-design of householder transformation,'' in \emph{29th International
  Conference on {VLSI} Design, {VLSID} 2016, Kolkata, India, January 4-8, 2016
  (in press)}.

\bibitem{fpu2}
F.~Merchant, N.~Choudhary, S.~K. Nandy, and R.~Narayan, ``Efficient realization
  of table look-up based double precision floating point arithmetic,'' in
  \emph{29th International Conference on {VLSI} Design, {VLSID} 2016, Kolkata,
  India, January 4-8, 2016}.

\bibitem{pipeline6}
\BIBentryALTinterwordspacing
M.~J. Flynn, P.~Hung, and K.~W. Rudd, ``Deep-submicron microprocessor design
  issues,'' \emph{IEEE Micro}, vol.~19, no.~4, pp. 11--22, Jul. 1999. [Online].
  Available: \url{http://dx.doi.org/10.1109/40.782563}
\BIBentrySTDinterwordspacing

\bibitem{pipeline5}
\BIBentryALTinterwordspacing
A.~Hartstein and T.~R. Puzak, ``The optimum pipeline depth for a
  microprocessor,'' in \emph{29th International Symposium on Computer
  Architecture {(ISCA} 2002), 25-29 May 2002, Anchorage, AK, {USA}}, 2002, pp.
  7--13. [Online]. Available: \url{http://dx.doi.org/10.1109/ISCA.2002.1003557}
\BIBentrySTDinterwordspacing

\bibitem{pipeline3}
\BIBentryALTinterwordspacing
------, ``Optimum power/performance pipeline depth,'' in \emph{Proceedings of
  the 36th Annual IEEE/ACM International Symposium on Microarchitecture}, ser.
  MICRO 36.\hskip 1em plus 0.5em minus 0.4em\relax Washington, DC, USA: IEEE
  Computer Society, 2003, pp. 117--. [Online]. Available:
  \url{http://dl.acm.org/citation.cfm?id=956417.956566}
\BIBentrySTDinterwordspacing

\bibitem{lac4}
\BIBentryALTinterwordspacing
A.~Pedram, A.~Gerstlauer, and R.~A. van~de Geijn, ``Floating point architecture
  extensions for optimized matrix factorization,'' in \emph{21st {IEEE}
  Symposium on Computer Arithmetic, {ARITH} 2013, Austin, TX, USA, April 7-10,
  2013}, 2013, pp. 49--58. [Online]. Available:
  \url{http://dx.doi.org/10.1109/ARITH.2013.21}
\BIBentrySTDinterwordspacing

\bibitem{laug}
E.~Anderson, Z.~Bai, C.~Bischof, S.~Blackford, J.~Demmel, J.~Dongarra,
  J.~Du~Croz, A.~Greenbaum, S.~Hammarling, A.~McKenney, and D.~Sorensen,
  \emph{{LAPACK} Users' Guide}, 3rd~ed.\hskip 1em plus 0.5em minus 0.4em\relax
  Philadelphia, PA: Society for Industrial and Applied Mathematics, 1999.

\bibitem{nick1}
\BIBentryALTinterwordspacing
N.~J. Higham, ``Exploiting fast matrix multiplication within the level 3
  {BLAS},'' \emph{{ACM} Trans. Math. Softw.}, vol.~16, no.~4, pp. 352--368,
  1990. [Online]. Available: \url{http://doi.acm.org/10.1145/98267.98290}
\BIBentrySTDinterwordspacing

\end{thebibliography}

\end{document}